\documentclass[twocolumn,showpacs,preprintnumbers,amsmath,amssymb]{revtex4}
\usepackage{epic,eepic,graphicx,epsf,axodraw,dcolumn,bm}
\newcommand{\beq}{\begin{equation}}
\newcommand{\eeq}{\end{equation}}
\def\bea{\begin{eqnarray}}
\def\eea{\end{eqnarray}}
\def\nn{\nonumber}
\def\la{\langle}
\def\ra{\rangle}

\begin{document}

\newcommand{\sheptitle}{Majorana Mass Zeroes from Triplet VEV without Majoron Problem}
\newcommand{\shepauthor}{P.H. Frampton, ~M.C. Oh ~and ~T. Yoshikawa}
\newcommand{\shepaddress}{Institute of Field Physics, Department of Physics and Astronomy, \\
University of North Carolina, Chapel Hill, NC 27599-3255.}

\title{\sheptitle}

\author{\shepauthor} 

\affiliation{\shepaddress}

\begin{abstract}
It is shown how to obtain recently-proposed two-zero
Majorana mass textures in models with three Higgs triplets
with small VEVs and a sufficiently massive triplet Majoron 
by using abelian discrete symmetries. It is briefly discussed
how in SU(5) grand unification where the triplets occur
in {\bf 15}'s the neutrino textures can be related to up-
and down- quark mass textures.
\end{abstract}

\maketitle

\section{Introduction}

As is well-known, in the standard model (SM) with only left-handed neutrino
fields $\nu_{iL}$, the neutrinos are necessarily massless because
a Majorana mass term $m_{ij} \nu_{iL}^{\alpha}\nu_{jL}^{\beta} \epsilon_{\alpha\beta}$
breaks $SU(2)$ gauge symmetry and is not renormalizable. Thus an extension
of the theory is necessary to accommodate the experimental observations
\cite{nuexpts, SNO}
of non-vanishing neutrino mass\cite{Barr}.

Keeping only left-handed neutrino fields for simplicity, there is an important
question of what is the most economical extension? One simple
possibility is surely the Zee model\cite{zee} which adds a singlet charged
scalar and results at one-loop order in a Majorana mass matrix
with vanishing diagonal entries. However, this model is now strongly
disfavored by the combination of SuperKamiokande and SNO data
\cite{FOY1, KOIDE}.

More generally, it was argued in \cite{FGM} that {\it any} Majorana 
matrix with three texture zeroes (including the Zee model) is
strongly disfavored phenomenologically. At most two such texture zeroes are
permitted and of the fifteen ways of assigning two zeroes
to the six inequivalen mass matrix entries
only seven (classified as A1, A2, B1, B2, B3, B4 and C in \cite{FGM})
survive comparison with the SuperKamiokande and SNO data.

Here we make an attempt to incorporate these permissable textures
into a model which contains the SM Higgs doublet (H) with $<H> \sim 100 $GeV
and
one or more Higgs triplets ($T_k$) (k = 1,2,...)
at least some having small non-vanishing VEV. In doing this, we must first address
the well-known problem\cite{majoron} of a triplet Majoron associated with making
a non-vanishing $<T> \neq 0$.

Next we incorporate the two-zero textures that are allowed phenomenologically.

Finally we discuss the interpretation in terms of $SU(5)$ unification
where the triplet $T_k$ fields appear in ${\bf 15}_k$ representations.

\section{ Triplet Higgs Model }

The triplet Higgs has Yukawa couplings only to the Majorana neutrinos. 
To get a sufficiently small neutrino mass from these interactions 
the vacuum expectation value (VEV) of the triplet Higgs has to be also 
very small to avoid incredibly small Yukawa couplings. 
The model with such triplet Higgs has a potential 
difficulty that spontaneous breakdown of lepton
number (L) will lead to a very light pseodoscalar 
Nambu-Goldstone boson ${\cal J}$ (Majoron) 
which does not agree with the experimental width of Z decay. 
Thus, our model must be such that this state ${\cal J}$ has mass
greater than half the $Z$ mass to avoid this Majoron problem.

For the triplet Higgs, ${\bf T} = \left( T^1, T^2,  T^3  \right)$ ,  
the Yukawa coupling to left-handed lepton doublet $L_L = (\nu, e^-)_L $ is 
\bea
{\cal L} &=& - f \overline{L^c_L} \sigma^a T^a L_L \nn \\
         &=& - f \overline{e^c} (T^1 - i T^2 ) e 
             + f \overline{\nu^c} (T^1 + i T^2 ) \nu \nn \\
         & & - f \overline{\nu^c} T^3 e - f \overline{e^c} T^3 \nu \\
         &=& - \sqrt{2} f \overline{e^c} T^{++} e 
             + \sqrt{2} f \overline{\nu^c} T^0 \nu \nn \\
         & & - f \overline{\nu^c} T^+ e - f \overline{e^c} T^+ \nu .
\eea
where
\bea
T^{++} &=& \frac{1}{\sqrt{2}}(T^1 - i T^2), \nn \\
T^{+} &=& T^3, \nn \\
T^{0} &=& \frac{1}{\sqrt{2}}(T^1 + i T^2) \nn .
\eea

The kinetic term of the triplet Higgs is 
\bea
\left| \partial_\mu {\mit T } - \frac{i}{2}  
            \left( \begin{array}{cc} 
                   \sqrt{g^2 + {g^\prime}^2} A_\mu &
                     \sqrt{2} g W^+ \\
                     \sqrt{2} g W^- & \sqrt{g^2 + {g^\prime}^2} Z_\mu \\
                   \end{array} \right) {\mit T } \right|^2,  
\eea
where
\bea
{\mit T } &\equiv& \sigma^a T^a \nn \\
             &=& \left( \begin{array}{cc}
                        T^+ & \sqrt{2} T^{++} \\
                       \sqrt{2} T^0 & - T^+ \\
                      \end{array} \right).  
\eea 
The $Z-T-T$ couplings are 
\bea
i \sqrt{g^2 + {g^\prime}^2} [ T^{0*} Z^\mu \partial_\mu T^0 
                            - \partial_\mu T^{0*} Z^\mu T^0 ].
\label{ZTT} 
\eea 

In this Majoron model, to give the tiny majorana neutrino mass 
$T^0 $ has to get a tiny VEV. The magnitude may be 
$\sqrt{\Delta_a} \sim O(0.1) eV$ if the Yukawa coupling is on order unity.
Then, we can decompose the neutral Higgs,  
\bea
T^0 = u + \frac{1}{\sqrt{2}}( \rho + i{\cal J}), 
\eea 
where $u$ is the VEV, $\rho$ is real part and ${\cal J}$ is the imaginary part. 
From 
eq.(\ref{ZTT}), the $Z-T-T$ couplings are 
\bea
\frac{1}{2} \sqrt{g^2 + {g^\prime}^2} Z^\mu [ (\partial_\mu \rho) {\cal J} 
                                            - \rho \partial_{\mu} {\cal J} ]. 
\eea
So on the $Z$ decay, it is $\sqrt{g^2 + {g^\prime}^2} m_z $. In this case, 
$Z$ boson can decay to $\rho $ and ${\cal J}$ because  
these masses are on order of $u$. If so, the ratio of the partial 
decay widths
\bea
\frac{\Gamma( Z\rightarrow \rho {\cal J}) }{\Gamma( Z\rightarrow \nu \nu ) } = 2, 
\eea
would mean an additional invisible width of Z. 
This is inconsistent with the experimental data where the
invisible width correspond quite precisely to that expected
for three active neutrinos and leaves no room for
triplet Majoron decay.   

To avoid this difficulty, we therefore need to extend the model. As 
methods to solve this difficulty, we can consider three cases.  
\begin{itemize}
\item[1)]{ By Higgs mechanism, $\rho $ and ${\cal J}$ have to be absorbed.} 
\item[2)]{ Giving a large mass to the Majorons without VEV. }
\item[3)]{Giving a large mass to the Majorons with tiny VEV.} 
\end{itemize}     

The first case will generally produce unacceptably-light gauge bosons. 

In the second case, we may consider a scenario in which 
neutrino masss arises from the radiative corrections. 
However, to permit a diagram contributing to Majorana
neutrino masses we need a trilinear coupling $\mu \phi T \tilde{\phi}$, 
where $\phi$ is the doublet higgs. When electroweak symmetry 
is broken, $\phi$ has a VEV and induces a shift 
in the VEV of $T$ as shown in Fig. {\bf 1}, because
the trilinear coupling contributes to a linear term, and
the small VEV for $T$ is thus destabilized.

Hence only scenario (3) is viable, and it requires that
$T$ has a very small VEV while ${\cal J}$, and $\rho$,
have heavy masses $M_{{\cal J},\rho} > M(Z)/2$.

\begin{center}
\setlength{\unitlength}{0.140900pt}
\begin{picture}(1500,900)(0,0)
\footnotesize
\thinlines \path(155,341)(1433,341)
\thinlines \path(794,135)(794,856)
\put(760,850){\makebox(0,0)[l]{\shortstack{$ V  $}}}
\put(1400,310){\makebox(0,0){$ T^0  $}}
\put(335,459){\makebox(0,0){$ \mu \la \phi \ra^2 T  $}}
\put(300,690){\makebox(0,0){$ M^2 T^2 + \lambda T^4  $}}
\put(1040,530){\makebox(0,0){$\bf --->   $}}
\thinlines \drawline[-95](342,856)(349,841)(362,813)(374,785)
(387,758)(400,732)(413,707)(426,682)(439,659)(452,636)(465,614)
(478,593)(491,573)(504,554)(516,535)(529,518)(542,501)(555,485)
(568,470)(581,455)(594,442)(607,429)(620,418)(633,407)(646,397)
(658,387)(671,379)(684,371)(697,365)(710,359)(723,354)(736,350)
(749,346)(762,344)(775,342)(788,341)(800,341)(813,342)(826,344)
(839,346)(852,350)(865,354)(878,359)(891,365)(904,371)(917,379)
(930,387)(942,397)(955,407)(968,418)(981,429)
\thinlines \drawline[-95](981,429)(994,442)(1007,455)(1020,470)
(1033,485)(1046,501)(1059,518)(1072,535)(1084,554)(1097,573)
(1110,593)(1123,614)(1136,636)(1149,659)(1162,682)(1175,707)
(1188,732)(1201,758)(1214,785)(1226,813)(1239,841)(1246,856)
\thicklines \path(415,856)(426,831)(439,802)(452,774)(465,747)
(478,721)(491,695)(504,671)(516,647)(529,624)(542,602)(555,581)
(568,561)(581,541)(594,523)(607,505)(620,488)(633,472)(646,456)
(658,442)(671,428)(684,416)(697,404)(710,393)(723,382)(736,373)
(749,364)(762,357)(775,350)(788,344)(800,339)(813,334)(826,331)
(839,328)(852,326)(865,325)(878,325)(891,326)(904,327)(917,330)
(930,333)(942,337)(955,342)(968,347)(981,354)(994,361)(1007,370)
(1020,379)(1033,389)(1046,399)(1059,411)
\thicklines \path(1059,411)(1072,423)(1084,437)(1097,451)(1110,466)
(1123,482)(1136,498)(1149,516)(1162,534)(1175,553)(1188,573)
(1201,594)(1214,616)(1226,638)(1239,662)(1252,686)(1265,711)
(1278,737)(1291,764)(1304,791)(1317,820)(1330,849)(1333,856)
\thinlines \drawline[-95](155,599)(155,599)(168,593)(181,588)
(194,583)(207,578)(220,572)(232,567)(245,562)(258,557)(271,552)
(284,546)(297,541)(310,536)(323,531)(336,526)(349,520)(362,515)
(374,510)(387,505)(400,500)(413,494)(426,489)(439,484)(452,479)
(465,474)(478,468)(491,463)(504,458)(516,453)(529,448)(542,442)
(555,437)(568,432)(581,427)(594,422)(607,416)(620,411)(633,406)
(646,401)(658,396)(671,390)(684,385)(697,380)(710,375)(723,370)
(736,364)(749,359)(762,354)(775,349)(788,344)
\thinlines \drawline[-95](788,344)(800,338)(813,333)(826,328)
(839,323)(852,318)(865,312)(878,307)(891,302)(904,297)(917,292)
(930,286)(942,281)(955,276)(968,271)(981,266)(994,260)(1007,255)
(1020,250)(1033,245)(1046,240)(1059,234)(1072,229)(1084,224)
(1097,219)(1110,214)(1123,208)(1136,203)(1149,198)(1162,193)
(1175,188)(1188,182)(1201,177)(1214,172)(1226,167)(1239,162)
(1252,156)(1265,151)(1278,146)(1291,141)(1304,136)(1305,135)
\end{picture}\\
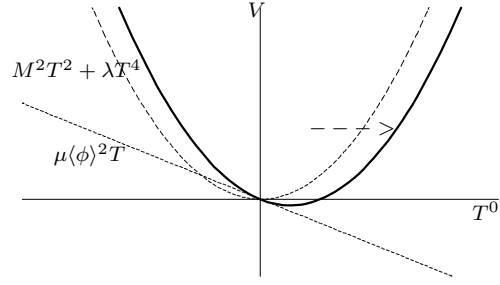
{\bf Fig. 1}
\end{center}

This third case which we employ
has been discussed by Ma and Sarkar\cite{masa}.
They induce the large mass of $\rho $ and ${\cal J}$ by adding 
lepton number violating terms in the Higgs potential. 

The most general Higgs potential of 
a doublet Higgs $\Phi = ( \phi^+, \phi^0 )$ and
a triplet Higgs $T = ( T^{++}, T^{+}, T^0 ) $ is  
\bea
V &=& m^2 \phi^\dagger \phi + M^2 T^\dagger T \nn \\
  &+& \frac{1}{2}\lambda_1 (\phi^\dagger \phi)^2 
   +  \frac{1}{2}\lambda_2 (T^\dagger T)^2 \nn \\
  &+& \lambda_3 (\phi^\dagger \phi)(T^\dagger T) \nn \\
  &-& \lambda_4 ( i \varepsilon_{ijk} \phi^\dagger \sigma^i \phi 
                         T^{\dagger j} T^k ) + h.c. \nn \\
  &+& \mu \phi^\dagger T^a \sigma^a \tilde{\phi} + h.c. .
\eea 
The important term is that with coupling $\mu $. Decomposing it, 
the terms are  
\bea
\mu \sqrt{2}[ \phi^0 \phi^0 T^{0*} 
              + \sqrt{2} \phi^+\phi^0T^- - \phi^+\phi^+T^{--} +h.c. ]
\nn  \\
\eea
The potential for neutral Higgses is 
\bea
&&m^2 \phi^{0*} \phi^0 + M^2 T^{0*}T^0 
          + \frac{\lambda_1}{2}(\phi^{0*} \phi^0 )^2 \nn \\
&+& \frac{\lambda_2}{2}(T^{0*} T^0 )^2 
          + (\lambda_3+\lambda_4)(\phi^{0*} \phi^0 )(T^{0*} T^0 ) \nn \\
&+& \sqrt{2} \mu ( \phi^0 \phi^0 T^{0*} 
                     + \phi^{0*} \phi^{0*} T^0 ) 
\eea

In this case, the conditions for stationarizing VEV are
\bea
m^2 + \lambda_1 v^2 + \lambda_3 u^2 + \lambda_4 u^2 + 2\sqrt{2}\mu u = 0 
\label{MINIMUM1}           \\
M^2 u + \lambda_2 u^3 + \lambda_3 v^2 u 
        + \lambda_4 v^2 u + \sqrt{2}\mu v^2 = 0  
\label{MINIMUM2}
\eea
The neutral Higgses are decomposed as follows:
\bea
T^0 &=& u + \frac{1}{\sqrt{2}}( \rho + i {\cal J}) \\
\phi^0 &=& v + \frac{1}{\sqrt{2}}( \xi_1 + i \xi_2 ) 
\eea 
To get the neutrino mass from the tiny VEV of the triplet Higgs, 
we need a hierarchy $ M, \mu \gg v \gg u $. When we assume the couplings 
$\lambda_i$  are parameters of order one, 
then, from eq. (\ref{MINIMUM2}),
\bea
u \sim - \frac{\sqrt{2} \mu v^2}{M^2}
\eea 
If we take $u \sim O(10^{-1})$ GeV, we need $M \sim 10^{14} GeV$ and $\mu $ is
also order $M$ or less.    
The mixing terms between the doublet one and the triplet one 
will appear by $\lambda_4 $ and $\mu$ terms. 
The mass matrices for  $(\xi_1, \rho)$  and $(\xi_2, {\cal J})$ are  
respectively 
\bea
\left( \begin{array}{cc}
      2 \lambda_1 v^2  & 2 (\lambda_3+\lambda_4) u v + 2 \sqrt{2}\mu v \\
      2 (\lambda_3+\lambda_4) u v + 2 \sqrt{2}\mu v & 
             2 \lambda_2 u^2 - \sqrt{2} \mu \frac{v^2}{u} \\
                      \end{array} \right)  \nn \\
\eea
and
\bea
\left( \begin{array}{cc}
      - 4\sqrt{2} \mu u &  2 \sqrt{2}\mu v \\
      2 \sqrt{2}\mu v  & 
             - \sqrt{2} \mu \frac{v^2}{u} \\
                      \end{array} \right)  
\eea
Thus, the mass eigenvalues are 
\bea
(\xi_1^m, \rho^m ) &\sim& ( 2 \lambda_1 v^2, - \sqrt{2} \mu \frac{v^2}{u} ) \\
(\xi_2^m, {\cal J}^m ) &=& ( 0, - \sqrt{2} \mu \frac{4 u^2 + v^2 }{u} )
\eea 
The would-be Nambu-Goldstone boson $\xi_2$ is eaten by $Z$ and the would-be 
Majoron ${\cal J}^m$ gets a large mass.  Because of the smallness of u, 
the mass of ${\cal J}^m$ will be much larger than $M_W$ if the parameter $\mu$ is
not so small value. Hence, in this case, $Z$ boson kinematically cannot decay to 
$\rho {\cal J}$ final state.   

The charged scalars also get the large mass. The masses of $T^{+m}$ 
and $T^{++}$ are proportional to 
$\mu \frac{v^2}{u} $. 

In this model, the radiative mass correction is 
\bea
\delta m_\nu \sim f \frac{m_l^2 \mu }{(4 \pi)^2 M^2 }
\eea
where $M$ is the charged triplet mass and about $10^{14}GeV$. This 
is at most $10^{-6}$ eV and so cannot be the main contribution 
to the neutrino mass, as it is too small to explain the LMA solution. 


\section{Incorporation of Two-Zero Textures.}

In \cite{FGM} it was pointed out that at most two
of the six independent elements of the symmetric
Majorana mass matrix can vanish and the phenomenologically-allowed
possibilities were classified as A1, A2, B1, B2, B3, B4, and C
respectively. It
is therefore of interest to accommodate these textures 
in our triplet Higgs model.
To do this, we impose on our model various discrete $Z_p$ symmetries 
which give rise to these two-zero textures
in a technically-natural manner.

\bigskip

The most realistic Majorana neutrino mass matrix to explain the 
experiments for solar and atmospheric neutrinos is 
\bea
\left( \begin{array}{ccc}
                        \delta & m_1 & m_2 \\
                       m_1 & \epsilon_1 & \epsilon_2 \\
                       m_2 & \epsilon_2 & \epsilon_3 \\
                      \end{array} \right). 
\eea
Here we need the mass hierarchy that is $ m_1 \sim m_2 
\gg \delta, \epsilon_i $ \cite{BabuMohapatra,FOY1}. This is corresponds
to Case C of the possible zeros textures in Ref.\cite{FGM}
if $\epsilon_1 = \epsilon_3 = 0$.   
For Case C \cite{FGM} we need at least three triplet Higgs $T_k$ (k = 1,2,3)
and we asign them under
$Z_3 $ as following, $\nu_1, \nu_2, \nu_3 \rightarrow \nu_1, \omega \nu_2, 
\omega^2 \nu_3$
and $T_1, T_2, T_3 \rightarrow T_1, \omega T_2, 
\omega^2 T_3$, where $\omega =\sqrt[3]{1}$ is a generator of $Z_3$. 
Then the Majorana neutrino mass matrix is
\bea
\left( \begin{array}{ccc}
    f_{11} \la T_1 \ra & f_{12}\la T_3\ra & f_{13}\la T_2\ra \\
    f_{12}\la T_3\ra & f_{22}\la T_2\ra & f_{23}\la T_1\ra \\
    f_{13}\la T_2\ra & f_{23}\la T_1\ra & f_{33}\la T_3\ra \\
                      \end{array} \right), 
\eea
where $\la T_i\ra$ is the VEV of the neutral triplet Higgs. 
Under a further $Z_2$, if $\nu_1 \rightarrow -\nu_1$, $\nu_2,\nu_3 \rightarrow 
\nu_2,\nu_3$, $T_1 \rightarrow T_1$ and $T_2,T_3 \rightarrow 
-T_2, -T_3$, The elements of $\{22\}$ and $\{33\}$ will disappear. And 
by taking $\la T_2\ra \sim \la T_3\ra > \la T_1\ra $, 
we can get most realistic 
mass matrix which is the C type texture. 
\bea
\left( \begin{array}{ccc}
 f_{11}\la T_1\ra & f_{12}\la T_3\ra & f_{13}\la T_2\ra \\
 f_{12}\la T_3\ra & 0 & f_{23}\la T_1\ra \\
 f_{13}\la T_2\ra & f_{23}\la T_1\ra & 0 \\
                      \end{array} \right).   
\eea

On the other hand, we can make also the Case $B1$ and Case $B2$.
By taking $\la T_2\ra = 0$ without $Z_2$ symmetry, 

\bea
\left( \begin{array}{ccc}
                       f_{11}\la T_1\ra & f_{12}\la T_3\ra & 0 \\
                       f_{12}\la T_3\ra & 0 & f_{23}\la T_1\ra \\
                       0 & f_{23}\la T_1\ra & f_{33}\la T_3\ra \\
                      \end{array} \right). 
\eea
For  $\la T_3\ra = 0$,  
\bea
\left( \begin{array}{ccc}
                       f_{11}\la T_1\ra & 0 & f_{13}\la T_2\ra \\
                       0 & f_{22}\la T_2\ra & f_{23}\la T_1\ra \\
                       f_{13}\la T_2\ra & f_{23}\la T_1\ra & 0 \\
                      \end{array} \right). 
\eea

\begin{table}
\begin{center}
\begin{tabular}{|c|c|c|c|c|c|c|c|}
\hline
   & 
$\nu_1$ & $\nu_2$ & $\nu_3$ & $T_1$ & $T_2$ & $T_3$ &  \\
\hline
$Z_3 \times Z_2$ & (1,-1) & ($\omega, 1) $ 
& ( $\omega^2, 1) $ & (1,1) & ($\omega $, -1) & ($\omega^2 $, -1) & C  \\
 \hline 
$Z_4 $ & 1 & i 
& -i  & 1 & i & -i & C  \\
 \hline
 \hline
$Z_4 \times Z_2$ & (1,1) & (i, -1)  
& ( -i, 1)  & (1,-1) & (-1, 1) & (i, 1) & $A_1$  \\
 \hline
$Z_4 \times Z_2$ & (1,1) & (-i, 1)  
& ( i, -1)  & (1,-1) & (-1, 1) & (i, 1) & $A_2$  \\
\hline
\hline
$Z_3$ & 1 & $\omega $ 
& $\omega^2 $ & 1 & $\times$ & $\omega^2$ & $B_1$  \\
 \hline
$Z_3$ & 1 & $\omega $ 
& $\omega^2 $ & $\times$ & $\omega $ & $\omega^2$ & $B_2$  \\
 \hline
$Z_4 \times Z_2$ & (i,-1) & (1, 1)  
& ( -i, 1)  & (1,-1) & (-1, 1) & (i, 1) & $B_3$  \\
\hline
$Z_4 \times Z_2$ & (i,-1) & (-i, 1)  
& ( 1, 1)  & (1,-1) & (-1, 1) & (i, 1) & $B_4$  \\
\hline
\end{tabular}
\caption{Accommodating two-zero textures of \cite{FGM}.} 
\end{center}
\end{table}

Proceeding along these lines, we can accommodate all the two-zero textures
of \cite{FGM} as indicated in the above Table
where the classification of Majorana neutrino mass matrices with two zeros in 
\cite{FGM} is used in the final column.

The triplet $T$ occurs naturally in the {\bf 15} of SU(5).
Although minmal $SU(5)$ is excluded, simple generalizations
are consistent with experimental constraints\cite{FG}.
Under $SU(5) \supset (SU(3)_C \times SU(2)_L)_Y$ the
various SU(5) irreps decompose as
\begin{eqnarray}
{\bf 5}  & \supset & (3, 1)_{-2/3} + (1, 2)_{+1} ~~~ \rm{(definition)} \nn \\
{\bf 15} \equiv ({\bf 5} \times {\bf 5})_s & \supset & (6, 1)_{-4/3} + 
(1, 3)_{+2}  + (3, 2)_{+1/3} \nn \\
{\bf \bar{15}} \equiv ({\bf \bar{5}} \times {\bf \bar{5}})_s & 
\supset & (\bar{6}, 1)_{+4/3} + (1, 3)_{-2} + (\bar{3}, 2)_{-1/3} \nn
\end{eqnarray}

\noindent and the $T$ triplet may be identified with the $(1, 3)_{\pm2}$ occurring
in ${\bf 15}, {\bf \bar{15}}$.

Let us give one example of how the neutrino Majorana mass texture may
be correlated with up- and down- quark
mass textures of the types discussed in {\it e.g.} \cite{RRR}.

Take the example in the above table with symmetry $Z_4 \times Z_2$
which gives the neutrino mass texture designated A2. By subsuming this in
$SU(5)$ with three {\bf 15}'s and requiring the three {\bf 10}'s
of fermions to transform as $(i,-1),(1,-1),(-1,1)$
and adding five {\bf 5}'s of Higgs transforming as
$(1,-1),(-i,-1),(i,1),(-i,1),(1,1)$.
gives one of the five permissable five-zero quark mass textures in \cite{RRR}.

We hope to return to a more detailed analysis of this neutrino-quark linkage
in a future publication.

\section*{Acknowledgments}
This work was supported in part by the US Department of Energy
under Grant No. DE-FG02-97ER-41036.

\end{document}